# High-performance Thermoelectric Monolayer γ-GeSe and its Group-IV Monochalcogenide Isostructural Family

Zheng Shu[a], Bowen Wang[a], Xiangyue Cui[a], Xuefei Yan[a], Hejin Yan[a], Huaxian Jia[b], Yongqing Cai[a,*]

[a]*Joint Key Laboratory of the Ministry of Education, Institute of Applied Physics and Materials Engineering, University of Macau, Taipa, Macau, China*

[b]*Beijing National Laboratory for Condensed Matter Physics and Institute of Physics, Chinese Academy of Sciences, Beijing 100190, China*

*Corresponding author: yongqingcai@um.edu.mo

## ABSTRACT

**Recently synthesized novel phase of germanium selenide (γ-GeSe) adopts a hexagonal lattice and a surprisingly high conductivity than graphite. This triggers great interests in exploring its potential for thermoelectric applications. Herein, we explored the thermoelectric performance of monolayer γ-GeSe and**

**other isostructural γ-phase of group-IV monochalcogenides γ-GeX (X = S, Se and Te) using the density functional theory and the Boltzmann transport theory. A superb thermoelectric performance of monolayer γ-GeSe is revealed with figure of merit *ZT* value up to 1.13-2.76 for n-type doping at a moderate carrier concentration of 4.73-2.58 × $10^{12}$ $cm^{-2}$ between 300 and 600 K. This superb performance is rooted in its rich pocket states and flat plateau levels around the electronic band edges, leading to promoted concentrations and electronic conductivity, and limited thermal conductivity. Our work suggests that monolayer γ-GeSe is a promising candidate for high performance medium-temperature thermoelectric applications.**

## 1. Introduction

Thermoelectric (TE) materials with high-efficiency, allowing a conversion of waste heat into electrical energy, potentially serve as components for alleviating key issues of energy storage and conversion in the 21st century. Nowadays, TE devices can achieve the reversible conversion between heat and electricity by the Seebeck and Peltier effects, boosting the application of power generation and electronic refrigeration [1-3]. For the TE devices, the efficiency of a power generation $\eta$ is determined by [1]

$$\eta = \frac{T_{\mathrm{H}}-T_{\mathrm{C}}}{T_{\mathrm{H}}}\left[\frac{\sqrt{1+ZT_{\mathrm{ave}}}-1}{\sqrt{1+ZT_{\mathrm{ave}}+\frac{T_{\mathrm{C}}}{T_{\mathrm{H}}}}}\right] \quad (1)$$

where $T_{\mathrm{H}}$ and $T_{\mathrm{C}}$ are the temperatures of hot and cold source respectively, and $ZT_{\mathrm{ave}}$ is an average value of $ZT$ from $T_{\mathrm{C}}$ to $T_{\mathrm{H}}$. Here, the $ZT$ is a dimensionless figure of merit defined as $ZT = \sigma S^2 T/(\kappa_{\mathrm{l}} + \kappa_{\mathrm{e}})$ and is used to characterize the TE performance at a specific temperature, where $\sigma$, $S$ and $T$ represent the electrical conductivity, the Seebeck coefficient and the absolute temperature, respectively. $\kappa_{\mathrm{l}}$ and $\kappa_{\mathrm{e}}$ are the contribution of the lattice and electronic components to the thermal conductivity ($\kappa$). A large $ZT$ value over 1 is essential for efficient conversion. According to this formula, two approaches can be used to improve the TE performance: (1) promoting the power factor ($\sigma S^2$) of materials; (2) reducing the $\kappa$ [4]. The goal of realizing a high-efficiency TE performance therefore is to find a suitable material with the increased $\sigma S^2$ and the reduced $\kappa$. However, it is a challenging task since the power factor and thermal conductivity are competing physical quantities [1].

In recent years, two-dimensional (2D) layered crystal materials have triggered great attention since the discovery of graphene [5]. Due to the electrons and phonons being confined to a limited space, 2D materials bring about opportunities to venture into the applications in nano electronic devices [6,7], resistive switching devices [8], advanced catalysts [9-11] and so on. Compared to their bulk counterparts, the 2D films made by nano-engineered technology can effectively avoid negative influences along the interlayer direction to reduce the lattice thermal conductivity [4,12,13]. Therefore, plenty of 2D materials were considered as promising thermoelectric materials [14-17] due to versatile electronic conductivities arising from varying density of states (DOS) near the Fermi level and modulated $\kappa_{\mathrm{l}}$. Exploring the physical mechanism of 2D materials for promoting thermoelectric performance is an ever-growing field.

Group-IV monchalcogenides MX (M = Ge, Sn; X = S, Se, Te), existing an isostructural lattice as phosphorene, have been a focus of commercial application for thermoelectric devices [18-24]. There are rich polymorphs for the monchalcogenides with the orthorhombic α- and β-phases being the normal phases. Zhao et al. reported bulk α-SnSe is a promising candidate for thermoelectric devices due to a very low lattice thermal conductivity [18,19]. P-type rhombohedral GeTe was shown to be an efficient thermoelectric material due to asymmetry-reduced lattice [20,21]. Hole doped orthorhombic GeSe was predicted to possess a high figure of merit $ZT$ of 2.5 at 800 K [22]. Furthermore, applying external pressure and doping are helpful means for promoting the thermoelectric performance of GeSe, which was validated by experiments [23,24].

Recently, a new phase of group-IV monochalcogenides with a hexagonal crystal structure, so-called monolayer γ-MX, has been theoretically predicted by first-principles calculations (DFT) [25]. The γ phase of germanium selenide (γ-GeSe), a member of the γ-MX family, has been successfully synthesized by chemical vapor deposition on the h-BN substrate [26]. Interestingly, γ-GeSe is a semi-metal in bulk phase and its measured average electronic conductivity in experiment is about $3\times10^5$ S/m [26], which is even higher than graphite. The electronic structures and optical properties of γ-MX were further explored by theoretical simulations [27-30]. Owing to the high conductivity together with a layered structure, it was demonstrated to have potential application in lithium-ion batteries with a small diffusion barrier of 0.21 eV for Li species [29]. As inspired by its superb electronic conductivity in bulk while semiconducting in monolayer [26], an intuition is that nanostructured γ-GeSe should have potential capability of conversion of thermal energy to electric energy. While group-IV monochalcogenides (α-phase) such as SnSe are renowned for a high electrical transport performance (power factor), the thermoelectric performance of this novel γ-phase is still a mystery.

In this work, we explored the novel γ-phase of the 2D group-IV monochalcogenides GeX (X = S, Se or Te) with respect to their thermoelectric applications by using first-principles calculations. By using the Boltzmann transport theory, together with the three-phonon scattering limited thermal conductivity, the figure-of-merit ZT values derived from electronic and thermal transport properties are predicted. We also pay attention to the lattice stability at various temperatures. Our work demonstrates that monolayer γ-GeSe is a superior thermoelectric material under n-type doping between 300 and 600 K. This can be attributed to an ultralow effective mass, high density of electronic states and a moderate lattice thermal conductivity. All these transport physical quantities can provide some hints for carriers dynamics for nanoelectronics and thermoelectrics.

## 2. Calculation methods

### *2.1. Density Functional Theory (DFT) Calculations*

All the DFT calculations involved in our work are implemented using the well-known Vienna *ab initio* simulation package (VASP) [33,34] within Projector Augmented-Wave (PAW) potentials. The generalized gradient approximation (GGA) of the Perdew-Burke-Ernzerhof (PBE) [35] functional is employed to describe the exchange-correlated interactions. The cut-off energy of plane wave is set to 500 eV. The monolayer GeXs are constructed by applying a vacuum slab of 15 Å to the X-Ge-Ge-X layer along the *Z*-direction. A Γ-centered 15 × 15 × 1 *k*-point grid is adopted for sampling the Brillouin zone. In addition, the convergence criteria of total energy and force acting on each atom is set to $1 \times 10^{-6}$ eV and 0.01 eV/Å, respectively. To obtain the accurate band structure, the Heyd-Scuseria-Elnzehof (HSE06) [36] hybrid functional is used to remedy the exchange-correlation interaction. A supercell of 5 × 5 × 1 (3 × 3 × 2) is used for monolayer (bulk) systems to obtain the phonon dispersion and the second-order harmonic interatomic force constants (IFCs) matrix based on density functional perturbation theory (DFPT) as implemented in Phonopy packages [37,38]. The *Ab initio* molecular dynamics (AIMD) simulations using a canonical ensemble (NVT) in the Nosé-Hoover heat bath [39] are carried out to investigate the thermal stability of the monolayer γ-GeX. In the supercell calculations for phonon dispersions and AIMD simulations, only the Γ point (0 0 0) in the reciprocal space is adopted.

## 2.2. Thermoelectric Properties Calculations

ShengBTE code [40] is used to calculate the $\kappa_l$ that demands the second-order harmonic and third-order anharmonic IFCs as input files. Based on the finite displacement method, the third-order anharmonic IFCs are calculated using 4 × 4 × 1 (3 × 3 × 2) supercell for monolayer (bulk) systems assisted with Thirdorder package. To acquire the accurate $\kappa_l$, a large cutoff distance up to 14 nearest atomic neighbors is used for the calculation of IFCs which was found to be enough [41], whereas 10 nearest atomic neighbors are considered for bulk system. Then, the tests of the number of Q-grid are performed to obtain the converged $\kappa_l$. Different from bulk materials, the calculated $\kappa_l$ of 2D materials should be normalized by multiplying a correction factor *l*/*d*, where *l* is the cell length along the *z*-direction and *d* is the thickness of the 2D layer.

In addition, the calculations of the electrical transport properties (Seeback coefficient *S* and the ratio of electrical conductivity to relaxation time σ/τ) are performed within BoltzTraP2 code [42]. The energy eigenvalues with a dense *k*-mesh of 30 × 30 × 1 is used to obtain the reliable electrical transport results. The calculations are based on constant relaxation time approximation. The relaxation time τ can be estimated by the deformation potential (DP) theory [43,44]:

$$\tau = \frac{\mu m^*}{e} = \frac{2\hbar^3 C}{3k_B T m^* E_1^2} \quad (2)$$

where $C$, $m^*$ and $E_1$ are the elastic constant, effective mass of electron or hole, and DP constant, respectively. These variables are described as the following equations

[43]:

$$C = [\partial^2 E/\partial(\Delta a/a_0)^2]/S_0 \qquad (3)$$

$$m^* = \hbar^2/(\partial^2 \varepsilon/\partial k^2) \qquad (4)$$

$$E_1 = \partial E_{\mathrm{edge}}/\partial(\Delta a/a_0) \qquad (5)$$

where $E$, $\varepsilon$, $k$ and $E_{\mathrm{edge}}$ are total energy of the system, the band energy, the electron wave vector and the energy of band edge under the bi-axial strain (-2%, -1%, 0, 1% and 2% as sampling points), $S_0$ is the area of an 2D material, and $\Delta a/a_0$ is the change ratio of lattice constant compared to the equilibrium state, respectively.

Electron-phonon coupling (EPC) theory is also carried out for the evaluation of relaxation time in the QUAMTUM ESPRESSO package [45]. The exchange-correlation of PBE with GGA is chosen to simulate the valence electrons, which is consistent with previous calculations implemented by VASP. The convergence threshold on total energy is $5 \times 10^{-6}$ Ry and the forces acting on each atom are smaller than $5 \times 10^{-6}$ Ry/a.u. with a kinetic energy cutoff of 70 Ry. The k-mesh of self-consistent calculation is 16 × 16 × 1 for monolayer γ-GeSe, and a q-mesh of 8 × 8 × 1 is used to calculate the phonon dispersion using DFPT method. To obtain the relaxation time, electronic self-energies Im ($\sum_{nk}$) are calculated using much denser k- and q-meshes of 160 × 160 × 1 and 80 × 80 × 1 within the electron-phonon Wannier (EPW) package [46].

## 3. Results and discussion

### *3.1. Crystal Structures and Electronic Properties of Monolayer γ-GeX*

The bulk γ-MX structures are stacked with γ-MX sheets bonding with weakly van der Waals interaction between X atoms. Each γ-MX sheet consists of four covalently bonded atomic sublayers (X-Ge-Ge-X) with an ABCA stacking sequence [24]. After full relaxation, the optimized geometrical structures of monolayer γ-GeX are illustrated in Fig. 1a and Fig. S1. The primitive cell contains four atoms with $P\bar{3}m1$ space group, where the Ge atoms occupy the sites at (0, 0) and ($a$/3, 2$b$/3), the X atoms are located at (2$a$/3, $b$/3). The optimized lattice constant of the monolayer γ-GeSe is $a = b = 3.76$ Å, which is in good agreement with previous theoretical reports [13-15]. Meanwhile, the bond lengths $l_{\text{Ge-Ge}}$ and $l_{\text{Ge-Se}}$ are 2.911 and 2.561 Å, respectively. In addition, the lattice constants of the monolayer γ-GeS and GeTe are $a$ = 3.61 and 4.01 Å, respectively. The detailed lattice parameters of three γ-GeX monolayers are shown in Table 1.

In order to assess the band structure accurately, the HSE06 [36] functional is carried out because PBE functional generally underestimates the band gap. Fig. 1b shows the comparison of band structures of monolayer γ-GeSe predicted by PBE and HSE06, respectively. In both cases, conduction band minimum (CBM) is located at the Γ point while the valence band maximum (VBM) is associated with two satellite states (at Γ-M and K- Γ sections respectively) aside the Γ point, leading to an indirect band gap of 0.623 eV (PBE) and 1.006 eV (HSE). Besides, we observe a broad conduction band extreme (CBE) along the Γ-M direction whose energy is slightly higher (0.016 eV) than that of CBM. The presence of such plateau and rich satellite states around the band edges induces high values of DOS which explains its high conductivity and are appealing for thermoelectric. The shapes of CBM and VBM calculated by HSE06 and PBE are nearly-identical. Similar tendency are also found in the monolayer γ-GeS and γ-GeTe (see Fig. S2a and S3a). The band gaps of three γ-GeX monolayers derived by HSE06 method are listed in Table 1.

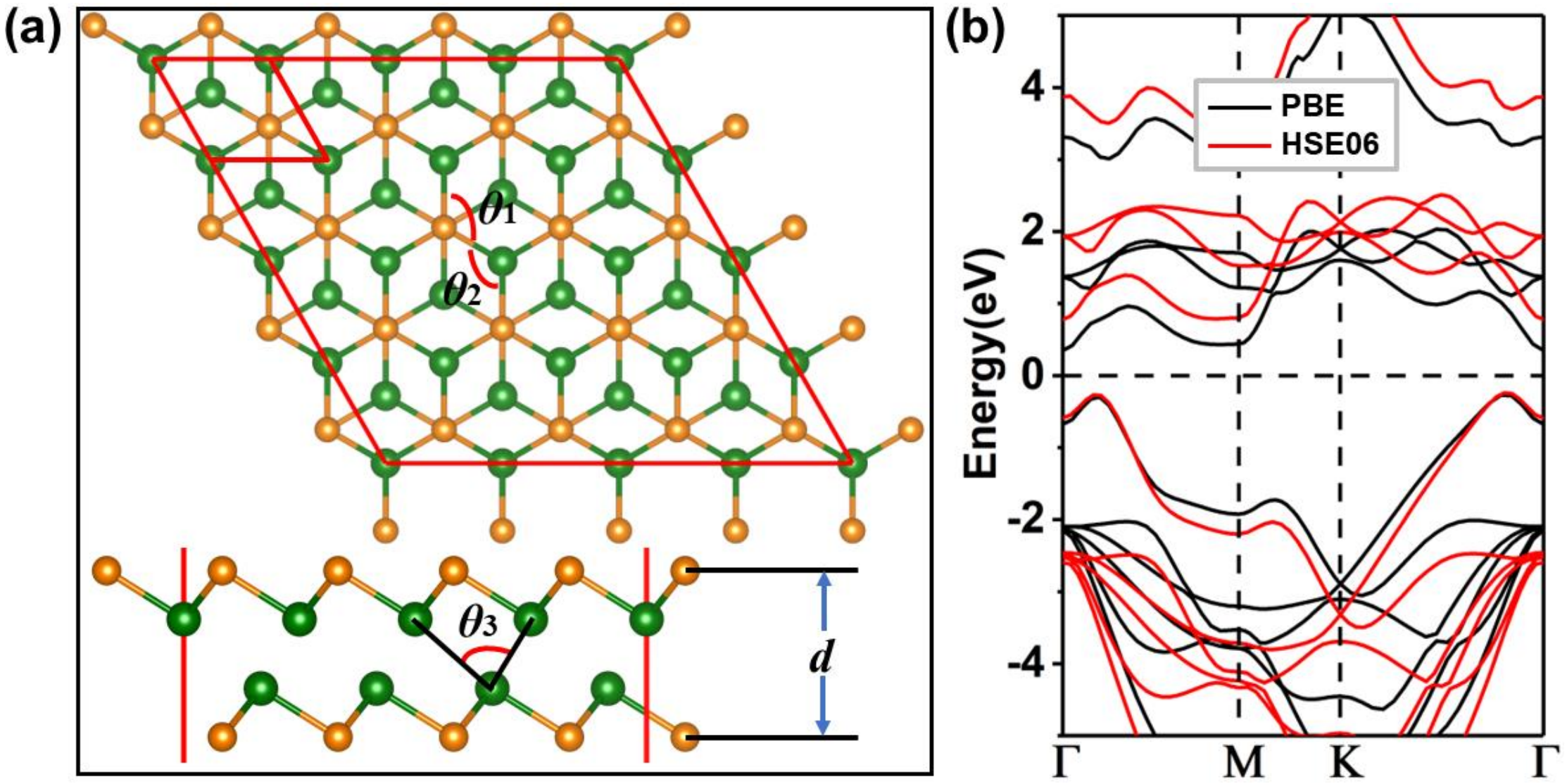

**Fig. 1.** The crystal structure of the (a) top view and (b) side view, and the (c) band structure of monolayer γ-GeSe. The orange and green balls represent Se and Ge atoms, respectively.

**Table 1.** Optimized Lattice Constants ($a$), Bond Length ($l_{Ge-Ge}$ and $l_{Ge-X}$), Bond Angles ($\theta_1$, $\theta_2$ and $\theta_3$) and Band Gap ($E_g$) with HSE06 functional for monolayer γ-GeX.

| | GeS | GeSe | GeTe |
|---|---|---|---|
| $a$ (Å) | 3.61 | 3.76 | 4.01 |
| $l_{Ge\text{-}Ge}$ (Å) | 2.895 | 2.911 | 2.941 |
| $l_{Ge\text{-}X}$ (Å) | 2.430 | 2.561 | 2.759 |
| $\theta_1$ (deg) | 95.95 | 94.44 | 93.23 |
| $\theta_2$ (deg) | 95.95 | 94.44 | 93.23 |
| $\theta_3$ (deg) | 77.15 | 80.45 | 85.96 |
| $E_g$ (eV) | 1.000 | 1.001 | 0.870 |

Meanwhile, the DFT-D3 scheme for the van der Waals (vdW) [47] correction and spin-orbit coupling (SOC) [48] effect are also examined. As shown in Fig. 2a-c, the band structures of monolayer γ-GeSe with vdW and SOC are similar to those without these corrections, especially for the CBM and VBM, suggesting the vdW and SOC have negligible effects on the electronic properties. Thus, subsequent calculations of electrical transport properties took no account of vdW and SOC effects. This is also true for the cases of γ-GeS and GeTe (see Fig. S2 and S3). The comparison of band gap with different corrections for γ-GeX is listed in Table S1. Our results are in agreement with the previous work [25]. Fig. 2d shows the density of states (DOS) of monolayer γ-GeSe. The large slope of DOS in electron pockets could increase the Seebeck coefficient for the n-type doping system.

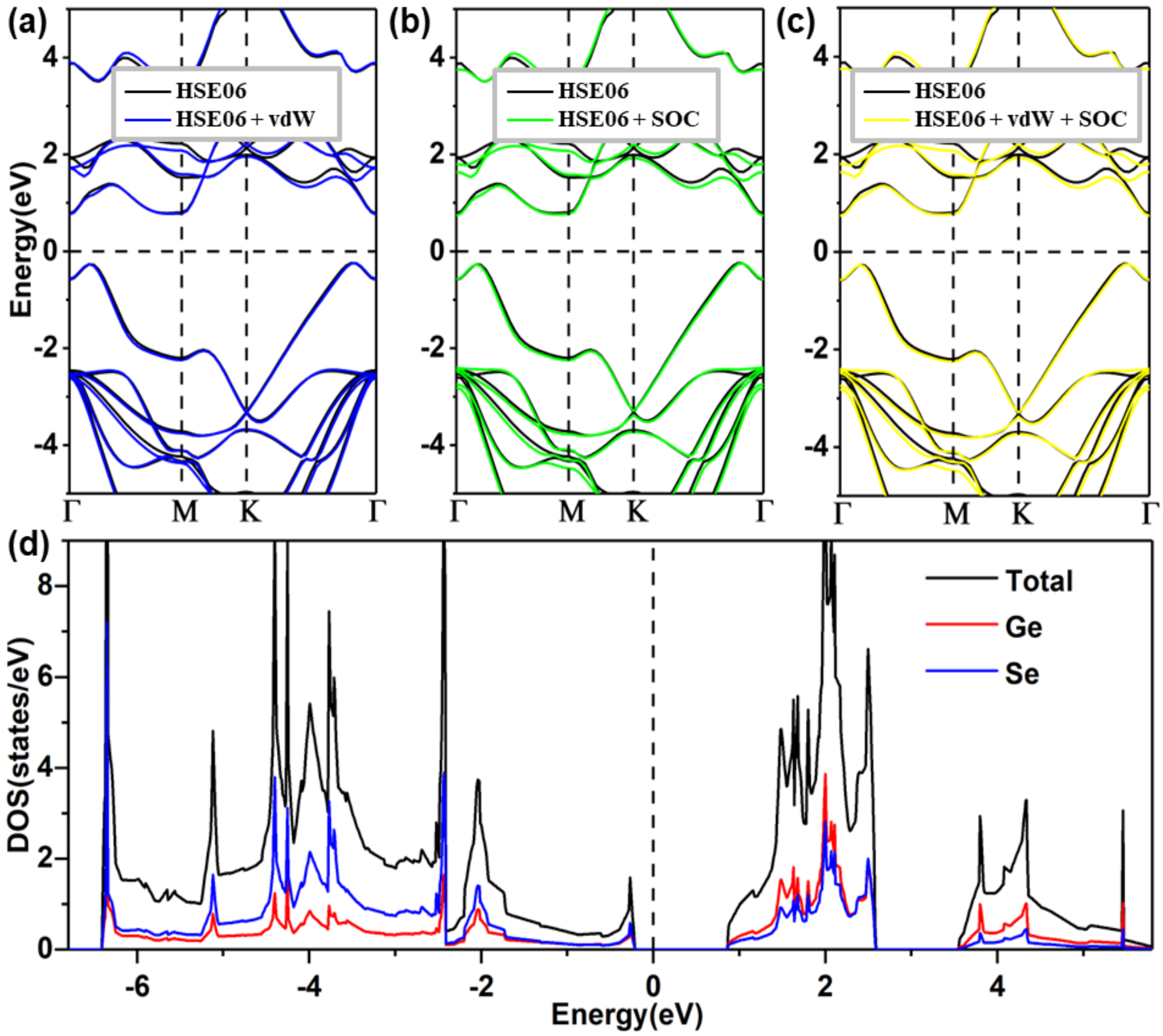

**Fig. 2.** (a-c) The effects of the SOC and vdW correction for the band structure and the (d) density of states (DOS) of monolayer γ-GeSe.

### *3.2. Dynamical and Thermodynamic Stabilities*

To assess the dynamical stability of monolayer γ-GeSe, the phonon spectrum is calculated as shown in Fig. 3a. With four atoms in the unit cell, there are 12 phonon modes with 3 acoustic modes and 9 optical modes. The acoustic modes consist of three lowest vibration frequency modes, which are the out-of-plane acoustic (ZA) mode, the transversal acoustic (TA) mode and longitudinal acoustic (LA) mode. The highest vibration frequency of monolayer γ-GeSe is 7.69 THz, which is much lower than γ-GeS (10.21 THz, see Fig. S4a). Moreover, we examine the dynamical structural stability under finite temperatures by using AIMD. Normally a robust and steady fluctuation of energy suggests a good thermodynamic stability. Herein we consider a temperature range of 100 to 600 K at which is normally the thermoelectric operating temperature. As revealed in Fig. 3b, the γ-GeSe maintains its stability up to 600 K while the structure collapses at 700 K.

The dynamical and thermodynamical stabilities of monolayer γ-GeS and GeTe are shown in Fig. S4-S6. It should be noted that imaginary frequency exists in the ZA mode of monolayer γ-GeTe, which is accordance with a previous report [25]. It is revealed that 2D monolayer of γ-GeTe may be not an exfoliable material from the 3D parent [49]. Based on this fact, the monolayer γ-GeTe is ruled out in the discussion of thermoelectric performance though it has a good thermal stability from a MD thermodynamic point of view (Fig. S6).

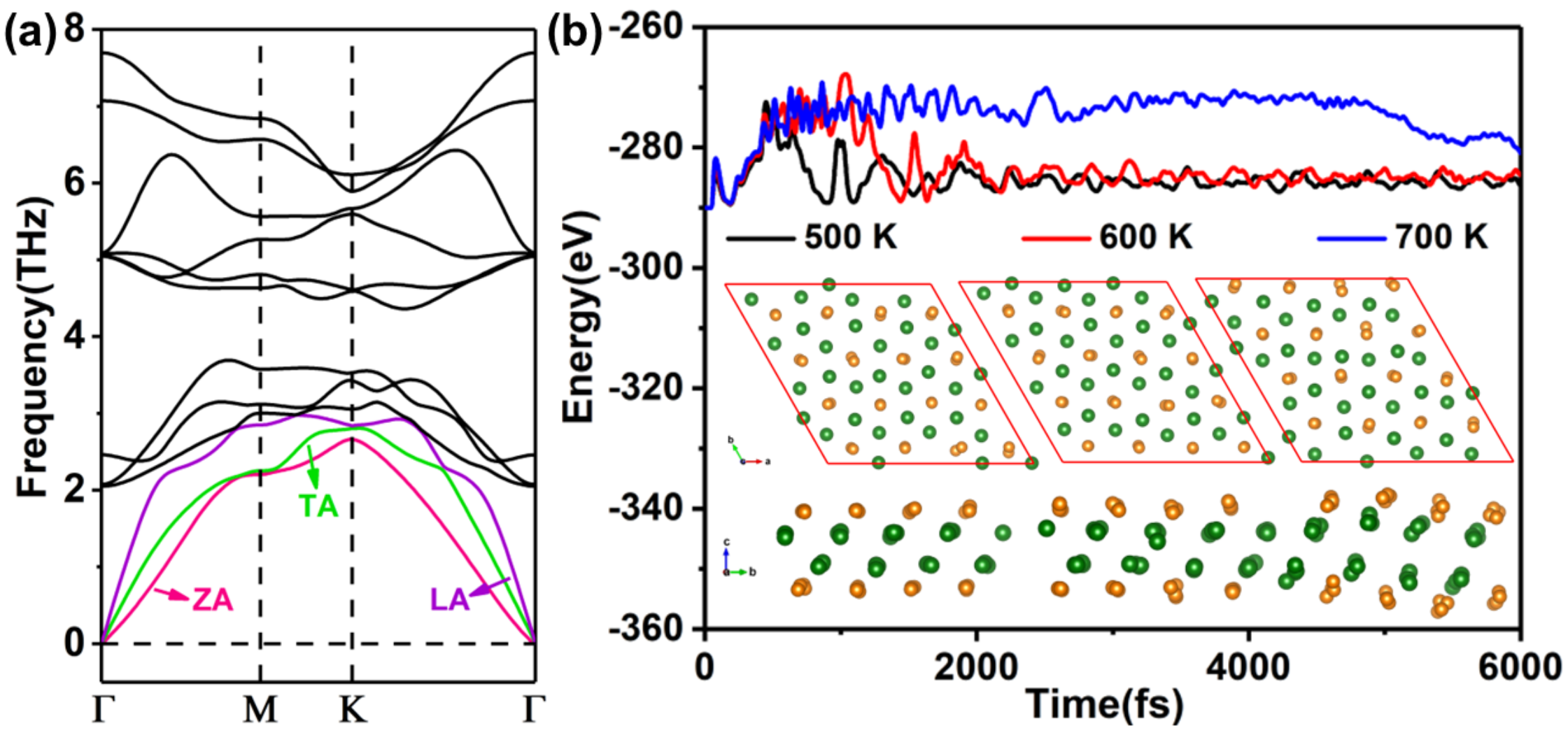

**Fig. 3.** (a) The harmonic phonon dispersion of monolayer γ-GeSe. (b) Fluctuations of energy with respect to time and equilibrium atomic structures of monolayer γ-GeSe at 500, 600 and 700 K.

### *3.3. Electronic Transport Properties*

The carrier concentration and relaxation time are important for determining the Seebeck coefficient, electrical conductivity and electronic thermal conductivity within the scheme of the Boltzmann transport theory. The Seebeck coefficient $S$ is independent on relaxation time, so that it can be directly derived as follows:

$$S = \frac{k_B}{e}\left[\ln(\frac{N}{n}) + 2.5 - r\right] \qquad (6)$$

where $e$, $N$, $n$ and $r$ are electron charge, effective density of states near fermi level, the number of carriers and scattering-mechanism parameter, respectively. Fig. 4a and b show the $S$ of monolayer γ-GeSe with p-type and n-type doping for a series of values of the carrier concentration in the temperature range of 300-600 K. As expected the $S$ varies reversely with carriers concentrations, and at 500 K the maximum value of $S$ is ~600 μV/K at carriers concentration of ~$10^{11}$ $cm^{-2}$. In contrast, the calculated $\sigma/\tau$ of monolayer γ-GeSe is positively proportional to the carriers concentration as plotted in Fig. 4c and d.

The relaxation time $\tau$ can be quantitatively estimated from Eq. (2) based on the DP theory, which is widely used to assess the performance of thermoelectric materials. To obtain $\tau$ on the basis of DP theory [42], the calculated elastic constant $C$, effective mass $m^*$ and DP constant $E_1$ are listed in Table S2 according to Eq. (3-5). In order to investigate the influence of chalcogen for the DP parameters, we also calculated that of monolayer γ-GeTe. As expected, a low effective mass can result in high thermoelectric performance [50]. More details about the calculations of relaxation time can be found in Supplementary data. Combined with all factors, the $\tau$ of monolayer γ-GeSe is found to be higher than those of GeS and GeTe, both for p-type or n-type doping. Fig. 5a shows the calculated $\sigma$ of monolayer γ-GeSe between 300 and 600 K which varies between $10^4$ and $10^6$ S/m and proportionally dependent on the carriers concentration.

Another key factor affecting the ZT is the thermal conductivity $\kappa$. The $\kappa$ consists of two sources: electrical part $\kappa_e$ and lattice part $\kappa_l$. According to Wiedemann-Franz law [51], the electronic thermal conductivity $\kappa_e$ can be estimated via:

$$\kappa_e = \kappa_0 - T\sigma S^2 = L\sigma T \qquad (7)$$

where $\kappa_0$ is the electrical thermal conductivity under closed-circuit condition, and $L$ is Lorenz constant, which is equal to 2.44 × $10^{-8}$ W·Ω·$K^{-2}$. The results of $\kappa_e$ using

Wiedemann-Franz law are proved to agree very well with the equation $\kappa_e = \kappa_0 - T\sigma S^2$ [52], therefore the $\kappa_e$ of monolayer γ-GeSe is obtained within Wiedemann-Franz law presented in Fig. 5b. The $\kappa_e$, reaching ~4 W/mK at $10^{13}$ $cm^{-2}$ for p dopants, nonlinearly increases with the carriers concentration while less depends on the temperature.

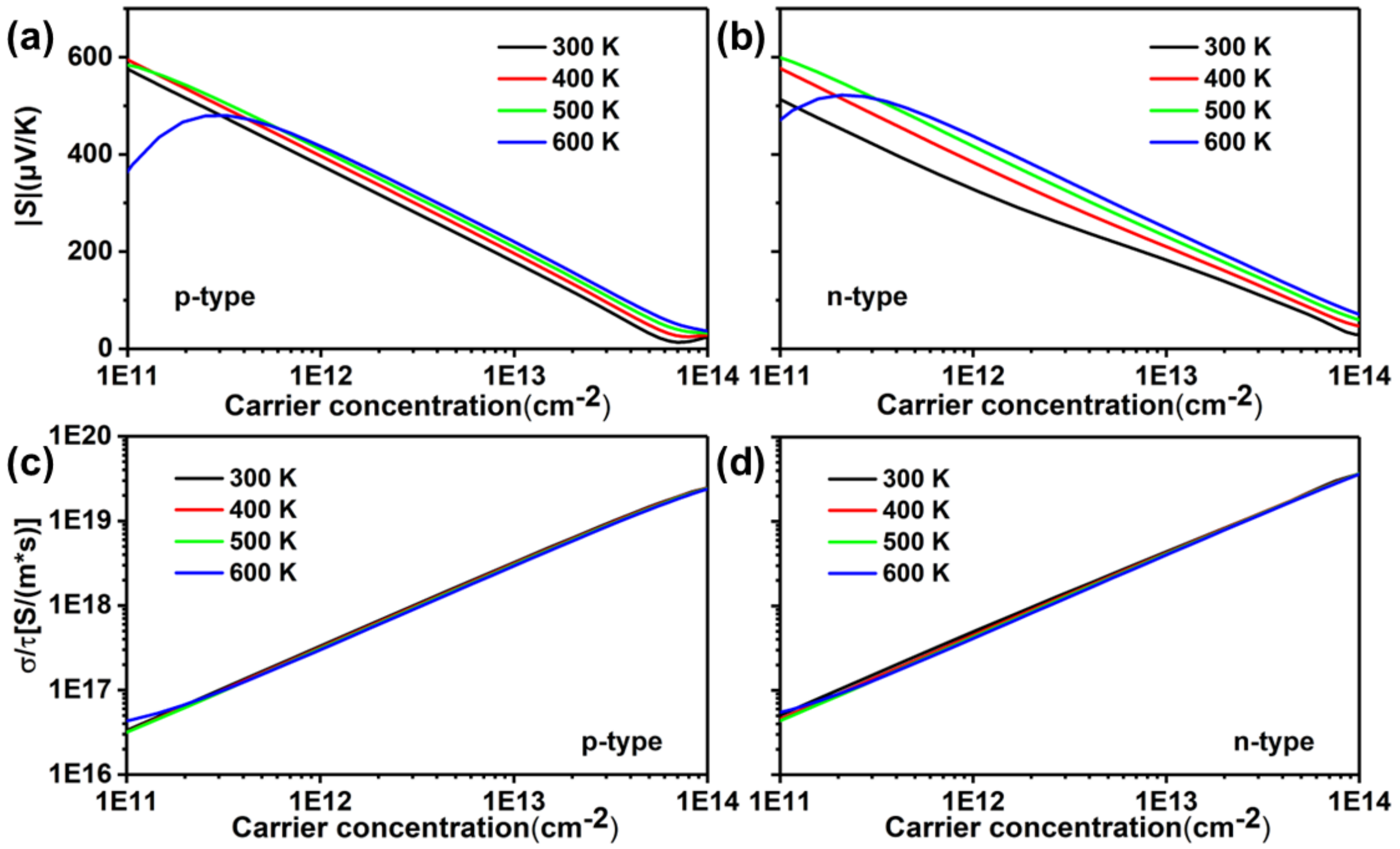

**Fig. 4.** (a) Seebeck coefficient $S$ and (b) the ratio of electrical conductivity to relaxation time σ/τ of monolayer γ-GeSe as a function of carrier concentration at 300, 400, 500 and 600 K.

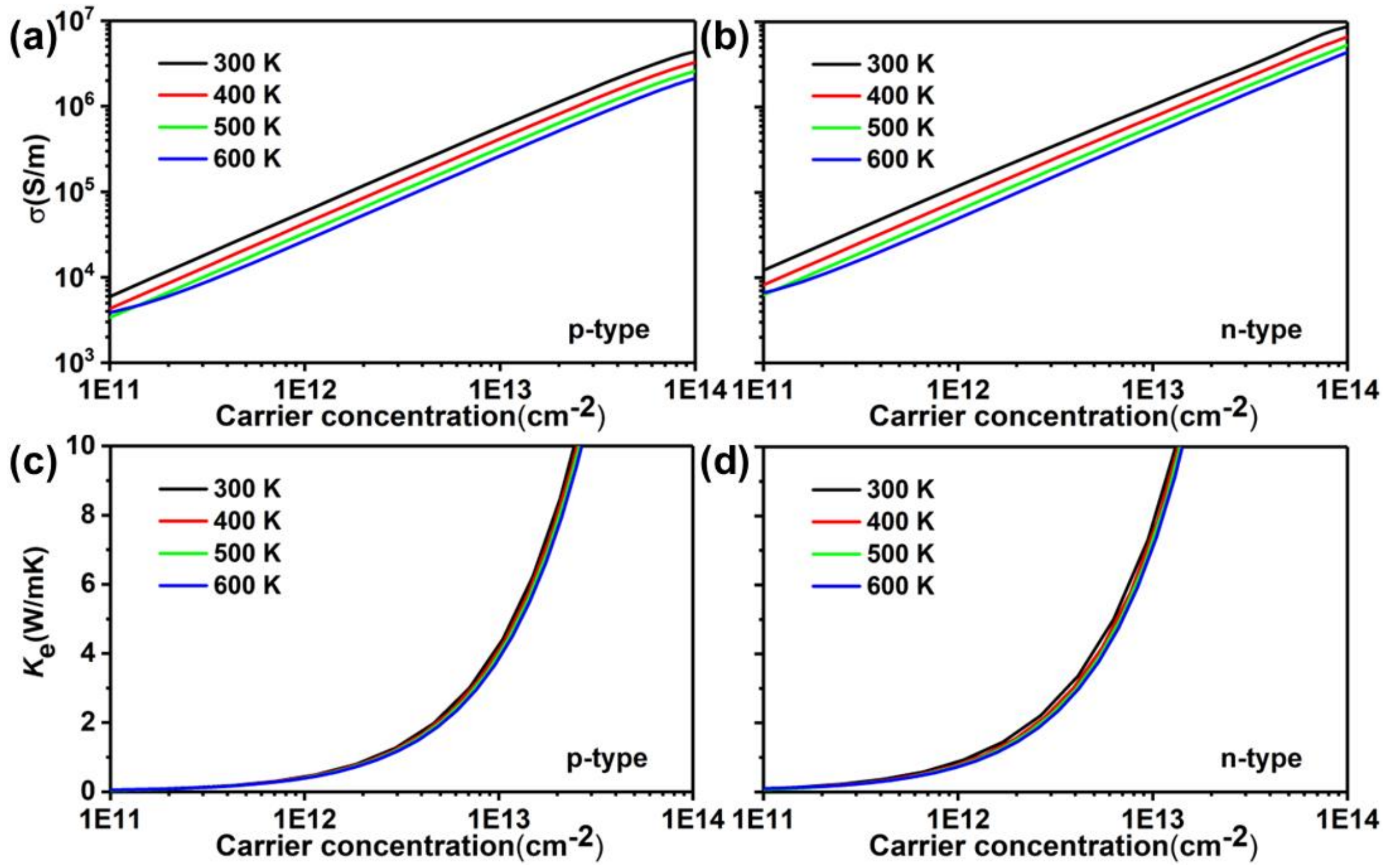

**Fig. 5.** (a) Electrical conductivity $\sigma$ and (b) electronic thermal conductivity $\kappa_e$ of monolayer γ-GeSe as a function of carrier concentration at 300, 400, 500 and 600 K.

### *3.4. Lattice Thermal Transport Properties*

The lattice thermal conductivity $\kappa_l$ is calculated through the phonon Boltzmann transport equation in ShengBTE [40] and expressed as

$$\kappa_l = \frac{1}{k_B T^2 N_0 \Omega} \sum_\lambda n_\lambda^0 (n_\lambda^0 + 1)\, v_\lambda^2 \tau_\lambda \omega_\lambda^2 \qquad (8)$$

where $N_0$ is the number of phonon wave vectors and $\Omega$ is the volume of the unit cell. Besides, $n_\lambda^0$, $v_\lambda$, $\tau_\lambda$, and $\omega_\lambda$ are the Bose-Einstein distribution function, the phonon group velocity, the relaxation time and the phonon angular frequency of phonon mode $\lambda$, respectively. As shown in Fig. 6a, the numerical values of $\kappa_l$ have minor fluctuations with Q. Here a 70 × 70 × 1 Q-grid is used to analyze the thermal transport properties.

The $\kappa_l$ decreases with increasing temperature owing to the promoted scattering cross section induced by stronger atomic vibrations at higher temperatures. The $\kappa_l$ at 600 K is 1.71 W/mK which is relatively low and desirable for thermoelectric applications.

To investigate the physical mechanism for the $\kappa_l$ of monolayer γ-GeSe, the anharmonicity of phonon is further discussed. Fig. 6b shows the mode distributed Gruneisen parameter γ of monolayer γ-GeSe, which can quantify the strength of phonon-phonon scattering. A large γ found for those low-frequency modes (ZA and TA) implies a strong anharmonicity and phonon scattering. As we can see, the large Gruneisen parameters (> 50) are associated with the ZA mode, suggesting that the low $\kappa_l$ at room temperature is mainly attributed to ZA acoustic mode. The phonon lifetime of monolayer γ-GeSe mainly lies between 0.1 and 10 ps which is shown in Fig. 6c. In this case, because the TA and LA modes have longer phonon lifetimes (> 10 ps) than ZA and optical branches, these two acoustic modes contribute more to $\kappa_l$ than others. According to Eq. (8), the $\kappa_l$ is strongly correlated with the phonon group velocity. The thermal energy is conveyed by the transport of phonons and the phonon group velocity is plotted in Fig. 6d where the zone-center LA branch has a higher velocity than TA, ZA and optical modes.

The $\kappa_l$ of monolayer γ-GeS is 1.07 W/mK at 300 K which is lower than γ-GeSe. This phenomenon is contrary to α-phase GeS and GeSe [22], implying a promising thermoelectric performance of γ-GeS at room temperature.

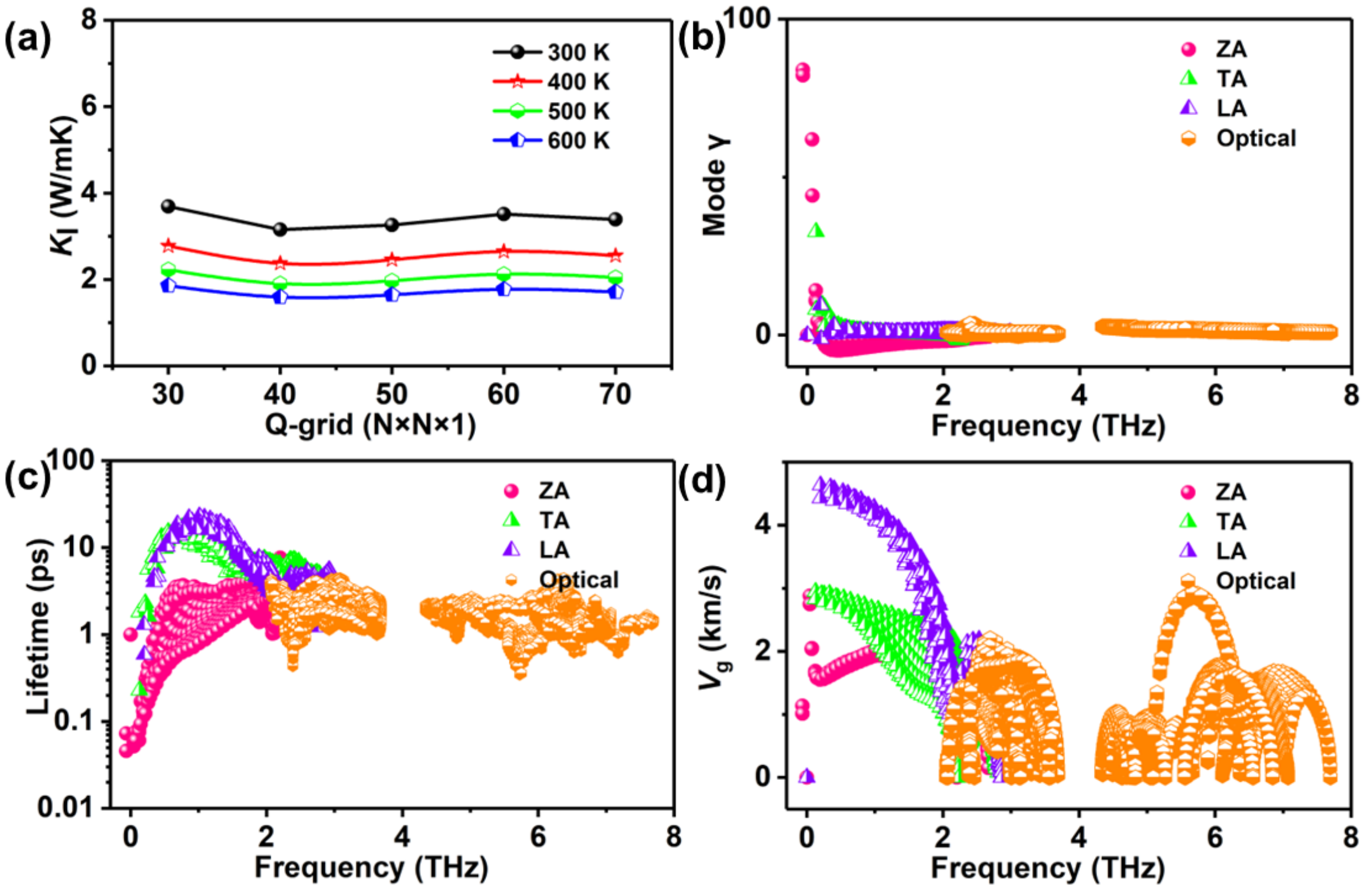

**Fig. 6.** (a) Lattice thermal conductivity $\kappa_l$, (b) Gruneisen parameter, (c) phonon group velocity and (d) phonon lifetime of monolayer γ-GeSe.

### *3.5. The Dimensionless Figure of Merit ZT value*

With the above analysis of electronic and thermal transport properties in mind, now we can quantitatively evaluate the thermoelectric performance. The figure of merits of *PF* and *ZT* of monolayer γ-GeSe as a function of carriers concentration are obtained and shown in Fig. 7. Owing to the compromise of the S and $\sigma$ under doping, the *PF* becomes maximum at ~$10^{13}$ cm$^{-2}$ with a slight shift for n and p doping cases. We found that n-type exhibits larger *PF* and *ZT* than p-type case, suggesting that n-type doping is more effective to enhance the thermoelectric performance of monolayer γ-GeSe than p-type doping. Furthermore, dimensionality reduction can significantly affect the transport of quantum quasi-particles and accordingly the thermoelectric performance. Therefore, we also calculated the $ZT$ of bulk γ-GeSe and compared the results with the monolayer phase. Different from the monolayer case, the relaxation time of bulk counterparts is obtained by:

$$\tau = \frac{2\sqrt{2\pi}\hbar^4 C}{3(k_B T m^*)^{3/2} E_1^2} \quad (9)$$

The effective mass $m^*$ of bulk materials is also different from the monolayer structure, which needs to consider the contribution of z-axis. The effective mass $m^*$ ($m_{xx}^*$, $m_{yy}^*$, $m_{zz}^*$) of bulk γ-GeSe is listed in Table S3. Based on DP theory and Eq. (9), the calculated phonon spectrum, band structure, lattice thermal conductivity and $ZT$ of bulk γ-GeSe are shown in Fig. 8. As we can see, the maximum $ZT$ along the in-plane direction increases from 0.14 to 0.35 as the temperature increases from 300 to 600 K for n-type doing. It is revealed that monolayer γ-GeSe has better thermoelectric performance than the bulk counterpart, indicative of the advantage of nano-engineering.

In addition, we also calculate the Seebeck coefficient $S$, electrical conductivity σ, power factor $PF$ and $ZT$ of monolayer γ-GeS at 300 K, which are shown in Fig. S7 and S8. Owing to a lower $\kappa_l$, the thermoelectric performance of monolayer γ-GeS is better than γ-GeSe at room temperature. The maximum $ZT$ value of monolayer γ-GeS at 300 K is pretty high (2.25 for n-type doping), implying a satisfying thermoelectric performance even at room temperature. For the comparison of the thermoelectric performance for several well-known thermoelectric materials, the scatter plot of $ZT$ value versus temperature of them is plotted and shown in Fig. 9. As we can see, our calculated $ZT$ value of monolayer γ-GeSe is high at medium temperature while that of γ-GeS functions well at room temperature.

The relaxation time calculated by DP theory does not consider the scattering of various phonons. Recent years, EPC [46] method can more accurately calculate the relaxation time [4,55,56]. Here, we recalculated the relaxation time of monolayer γ-GeSe for n-type doping and compared the results and $ZT$ with that of DP theory. In Fig. S10a, we plotted the energy-dependent relaxation time at 600 K. It is found that monolayer γ-GeSe exhibits higher n-type carrier relaxation time than p-type, which is consistent with the results calculated by DP theory. The comparison of n-type carrier relaxation time $\tau$ and $ZT$ using DP and EPC method is shown in Table 2 and Fig. S10b. As we can see, the $ZT$ calculated with EPC method is smaller than that of DP theory. However, the $ZT_{\text{max}}$ is still over 2 for n-type monolayer γ-GeSe using EPC method. In addition, the calculation of relaxation time by EPC is time- and memory-consuming and beyond the current computational capacities for more complex materials. Therefore, DP theory is overall a viable method to reasonably obtain the relaxation time.

**Table 2.** The comparison of relaxation time $\tau$ using DP and EPC method for n-type monolayer γ-GeSe at 300, 400, 500 and 600 K.

| | DP theory | | EPC theory | |
|---|---|---|---|---|
| | $\tau$ (fs) | $ZT_{max}$ | $\tau$ (fs) | $ZT_{max}$ |
| 300 K | 240.82 | 1.13 | 140.86 | 0.84 |
| 400 K | 180.62 | 1.72 | 101.32 | 1.26 |
| 500 K | 144.49 | 2.27 | 79.43 | 1.66 |
| 600 K | 120.41 | 2.76 | 65.45 | 2.04 |

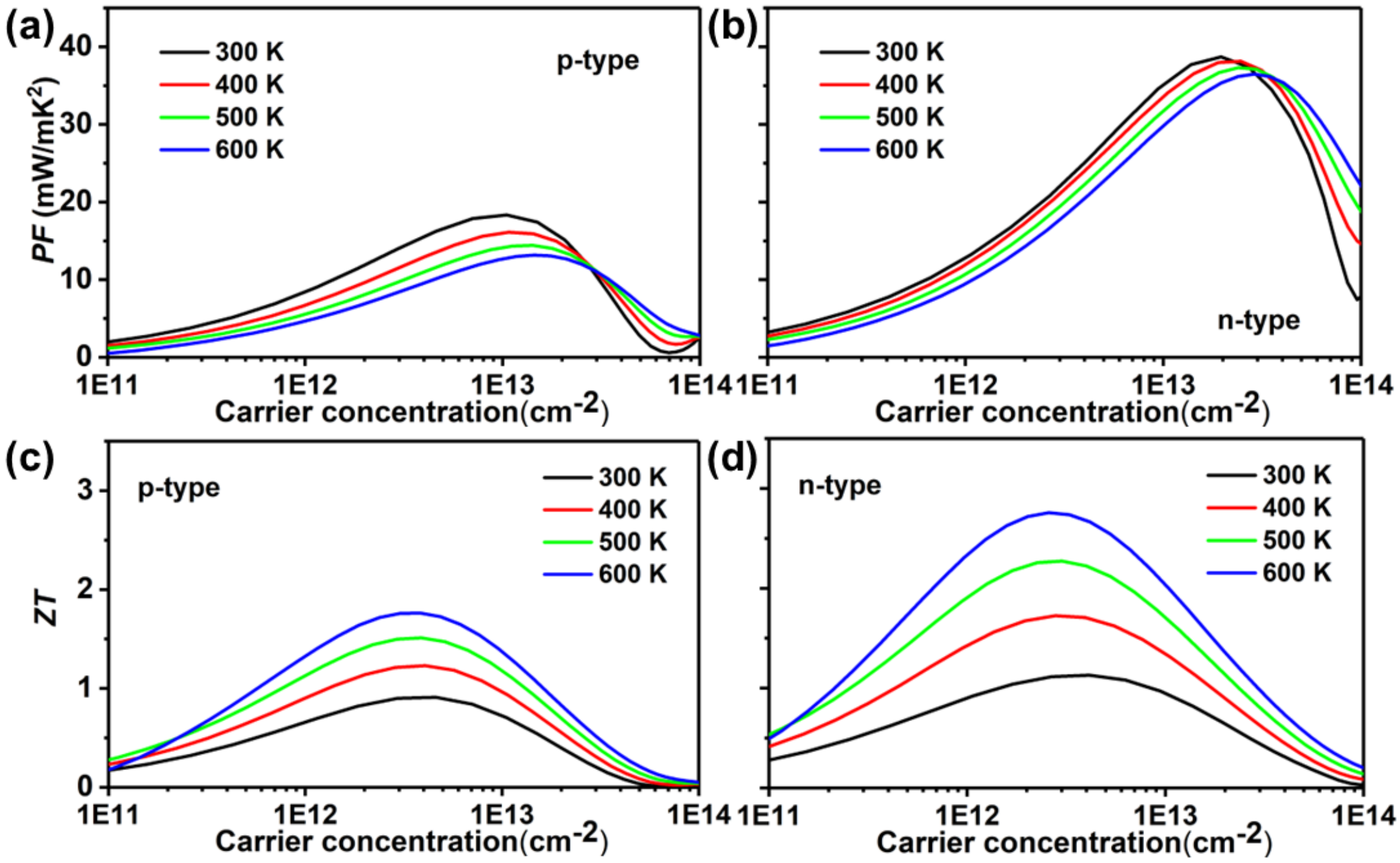

**Fig. 7.** (a) The power factor $PF$ and the (b) figure of merit $ZT$ of monolayer γ-GeSe as a function of carrier concentration at 300, 400, 500 and 600 K.

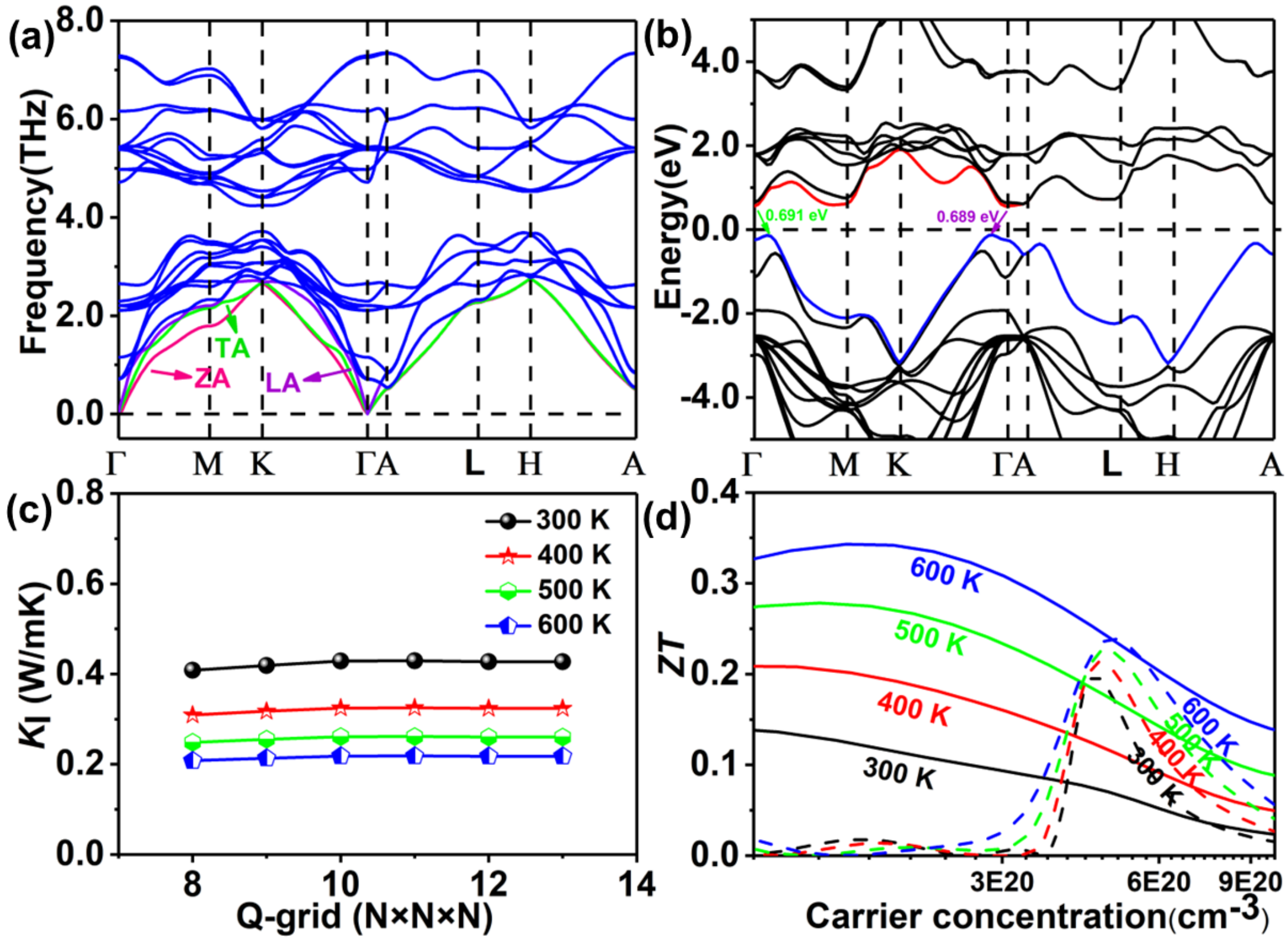

**Fig. 8.** (a) The harmonic phonon dispersion, (b) band structure (HSE06), (c) Lattice thermal conductivity $\kappa_l$ and the (d) carrier concentration-dependent figure of merit $ZT$ of bulk γ-GeSe. The $\kappa_l$ and $ZT$ are along the in-plane direction: solid and dashed lines are plotted for n- and p-type doping, respectively.

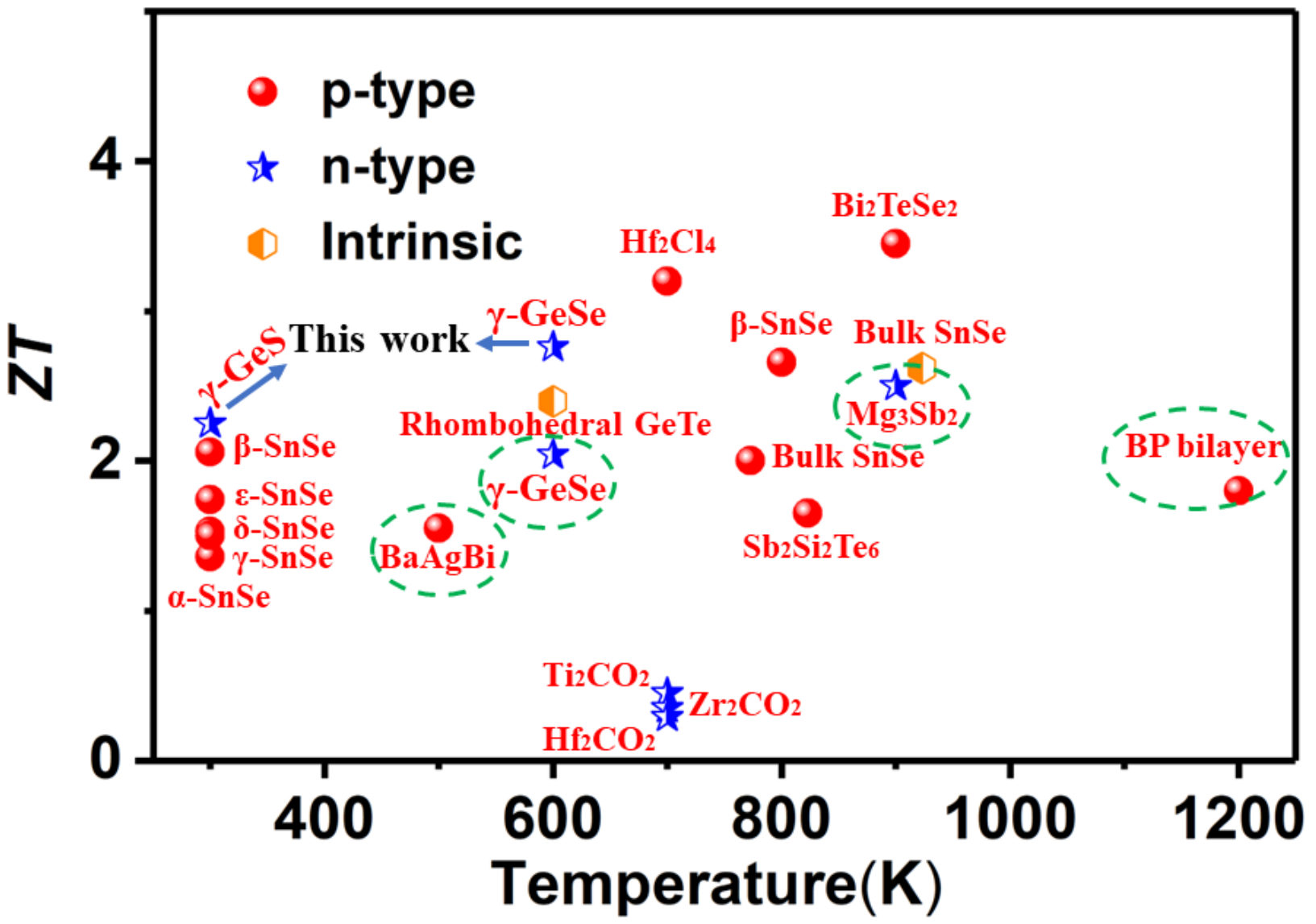

**Fig. 9.** The comparison of the thermoelectric performance of monolayer γ-GeSe with some selected thermoelectric materials [4,15-17,21,22,53-56]. The ZT values calculated by EPC method are marked by dashed line.

## 4. Conclusion

In summary, for the first time we perform systematic first-principles calculations with Boltzmann transport equations to assess the thermoelectric performance of monolayer γ-GeX. We find that even monolayer γ-GeSe possesses an intriguingly high electronic conductivity and simultaneously a poor thermal conductivity. These could be dated back to the high DOS associated with the valley states and flat bands around the band edges together with a relatively softening lattice. Our results show the highest figure of merit *ZT* value up to 1.13 at 300 K and 2.76 at 600 K for n-type doping in the case of monolayer γ-GeSe compared with other γ-GeX. In addition, dimensionality reduction can significantly improve *ZT* of γ-GeSe. As a thumb of rule for exploring high-performance thermoelectric materials, the presence of relatively flat bands around the band edges, a narrow band gap, rich-distributed pocket electronic states, as found in the layered γ-GeSe, is beneficial for high-density of states, efficient degenerate doping and accordingly a high power factor. Provided with a soft vibrating bond with a long radius of cationic and anionic atoms, a strong anharmonicity would generally be expected and a low thermal conductivity would be guaranteed, thus promoting the ZT value. We expect that this work could pave a way for the discovery of novel thermoelectric materials based on γ-phase group-IV monochalcogenides for thermoelectric applications.

## Declaration of Competing Interest

The authors declare that they have no known competing financial interests or personal relationships that could have appeared to influence the work reported in this paper.

## Acknowledgements

This work is supported by the Natural Science Foundation of China (Grant 22022309) and Natural Science Foundation of Guangdong Province, China (2021A1515010024), the University of Macau (SRG2019-00179-IAPME, MYRG2020-00075-IAPME) and the Science and Technology Development Fund from Macau SAR (FDCT-0163/2019/A3). This work was performed in part at the High-Performance Computing Cluster (HPCC) which is supported by Information and Communication Technology Office (ICTO) of the University of Macau.